\documentclass[twocolumn,showpacs,pre,superscriptaddress]{revtex4}
\usepackage{graphicx}
\usepackage{color}
\usepackage{amssymb,amsfonts,amsmath}
\usepackage{natbib}
\usepackage{ulem}

\begin{document}

\title{Benjamin-Feir instabilities on directed networks}
\author{Francesca Di Patti} \affiliation{Dipartimento di Fisica e Astronomia, Universit\`{a} degli Studi di Firenze,  CSDC and INFN, via G. Sansone 1, 50019 Sesto Fiorentino, Italia}
\author{Duccio Fanelli} \affiliation{Dipartimento di Fisica e Astronomia, Universit\`{a} degli Studi di Firenze,  CSDC and INFN, via G. Sansone 1, 50019 Sesto Fiorentino, Italia}
\author{Filippo Miele} \affiliation{Dipartimento di Fisica e Astronomia, Universit\`{a} degli Studi di Firenze, via G. Sansone 1, 50019 Sesto Fiorentino, Italia}
\author{Timoteo Carletti}\affiliation{Department of Mathematics and Namur Center for Complex Systems - naXys, University of Namur, rempart de la Vierge 8, B 5000 Namur, Belgium}

\date{\today} 

\begin{abstract}
The Complex Ginzburg-Landau equation is studied assuming a directed network of coupled oscillators. The asymmetry makes the spectrum of the Laplacian operator complex, and it is ultimately responsible for the onset of a generalized class of topological instability, reminiscent of the Benjamin-Feir type. The analysis is initially carried out for a specific class of networks, characterized by a circulant adjacency matrix. This allows us to delineate analytically the domain in the parameter space for which the generalized instability occurs. We then move forward to considering the family of non linear oscillators coupled via a generic direct, though balanced, graph.  The characteristics of the emerging patterns are discussed within a self-consistent theoretical framework. 
\end{abstract}


\maketitle
\section{Introduction}
Spatially extended systems can organize in space and time yielding collective macroscopic patterns \cite{pikovsky,turing,murray2,maini} . In many cases of interest, self-organization follows a spontaneous symmetry breaking instability triggered by random fluctuations of a uniform equilibrium \cite{mimura, baurmann,alan2,deanna, zhab,fanelli,biancalani,dipatti}.  The imposed perturbation seeds a resonant amplification mechanism that can eventually crystallize in striking motifs, a coordinated patchy-like assembly of shapes and forms. Patterns are widespread in nature \cite{cross}. For this reason, it is of paramount importance to systematically elaborate on the key ingredients which fuel the propensity of many body systems towards unsupervised pattern formation. 

Modulational instability occurs for instance when deviations from a periodic waveform are sustained by nonlinearities, leading to spectral-sidebands and the breakup of the waveform into a train of pulses. This is the so called Benjamin-Feir (BF) instability \cite{benjamin}, named after the researchers who first identified the phenomenon working with periodic surface gravity waves (Stokes waves) on deep water.  The BF instability is often illustrated in the framework of the Complex Ginzburg-Landau equation (CGLE) \cite{diprima, kuramotoBook}, a paradigmatic model for non linear phenomena, whose applications range from superconductivity, superfluidity and Bose-Einstein condensation to liquid crystals and strings in field theory.  

In a recent paper \cite{dipatti2016}, we revisited the BF instability, by operating with the CGLE modified with the inclusion of a drift term. A generalized class of BF instabilities was shown to occur instigated by the drift, and outside the region of parameters for which the conventional BF instability is expected to manifest.  These conclusions have been reached for a CGLE embedded in a continuum, one dimensional, spatial support.  In many cases of interest  it is however more appropriate to consider a finite population of discrete non linear oscillators \cite{strogatz1,strogatz2}. These latter occupy the nodes of a network \cite{mik,gwe,kouvaris,arenas,massaro,dedomenico},  links pairing interacting oscillators. Particularly relevant for our purposes is the case of an asymmetric network of connections. When reaction-diffusion systems are placed on directed, hence asymmetric graphs, patterns can arise, even if they are formally impeded on a  continuum domain or on discrete spatial medium \cite{asllani14, contemori}. Directionality induces an effective degree of spatial anisotropy, analogous to drift in the continuum limit, which can actively promote a pattern forming derive.  The conditions for the onset of an asymmetry driven instability, reminiscent of a Turing like mechanism,  for a multi-species reaction diffusion model constrained to evolve on a directed graph have been worked out in \cite{asllani14}. In this case, the perturbation is acted on a homogeneous fixed point, namely a time independent equilibrium for the reactions terms. 

Starting from this setting, the aim of this paper is to revisit the analysis in \cite{dipatti2016} for a family of non linear coupled oscillators, hosted on a directed network.  To this end, we initially specialize on a peculiar class of networks, characterized by a circulant adjacency matrix.  This choice enables us to progress analytically in delineating the region of the relevant parameters space for which the generalized BF instability is predicted to hold. As in \cite{dipatti2016} the homogeneous limit cycle, perturbed by a tiny externally imposed disturbance, 
can give rise to traveling waves or patchy intermittent patterns, which share a striking similarity with those obtained in the region classically deputed to the BF instability. In the second part of the paper we consider instead an ensemble made of non linear Ginzburg-Landau  oscillators, coupled via a generic direct graph. The conditions for the asymmetry driven BF instability are worked out by generalizing the approach of \cite{asllani14} to a setting where the homogeneous solution depends explicitly on time.  As we shall clarify in the following, by tuning the topology of the network of connections and consequently changing the spectrum of the associated Laplacian operator, can prompt an instability of the generalized BF class.

The paper is organized as follows. In the next Section we will introduce the reference model. In Section \ref{circulant} we shall consider the CGLE defined on a periodic lattice, modified with the inclusion of asymmetric long-range (next-nearest neighbors) links. Working in this setting we will determine the conditions that control the outbreak of the generalized BF instability. The phenomenon is also substantiated numerically. In Section \ref{network} we will turn to considering the case of a generic, though balanced, directed network. Finally, in the last Section, we will sum up and draw conclusions. 
\section{The model} \label{model}
Consider an ensemble made of $N$ non linear oscillators and label with  $W_j$ their associated complex amplitude, where $j=1,...,N$.  We shall hereafter assume that each oscillator obeys to a CGLE. Oscillators are placed on the nodes of a network: their mutual interaction is specified by a discrete Laplacian operator. Assume $\mathbf{A}$ to label the (potentially weighted) adjacency matrix of the network of contacts and denote by $k_j=\sum_kA_{jk}$ the connectivity of generic node $j$. Then, the entries of the associated Laplacian matrix $\mathbf{\Delta}$ are defined as $\Delta_{jk}=A_{jk}-k_j \delta_{jk}$.  The spatially extended CGLE that we shall hereafter analyze can be hence cast in the form: 
\begin{equation}\label{eq:GLnetwork}
\frac{d}{d t} W_j =  W_j - (1+i c_2) \vert W_j \vert ^2 W_j + (1+i c_1) \sum_k \Delta_{jk} W_k
\end{equation}
with $c_1$ and $c_2$ real parameters. In the following, we will solely consider perfectly balanced networks, namely, graphs characterized by an identical number of ingoing and outgoing links. The above Eq. (\ref{eq:GLnetwork}) admits therefore an homogeneous limit cycle solution of the form $W_j \equiv W^{LC} =\exp(-i c_2 t)$, $\forall j$. This latter is here termed the limit cycle (LC) solution, as it results from a uniform, fully synchronized, replica of the periodic orbit displayed by the system in its a-spatial version. Under specific conditions an external non homogeneous disturbance superposed to the initially synchronized configuration, can turn unstable. The imposed perturbation grows in time and the synchronization is eventually lost. Non homogeneous patchy like patterns emerge in the non linear regime of the evolution and represent the late time imprint of the so called Benjamin-Feir (BF) instability to which we alluded in the introductory paragraph.  When the CGL dynamics is hosted on a symmetric graph (e. g. a lattice), the BF instability sets in provided $1+c_1 c_2 <0$, a condition that we shall obtain as a particular limiting case of our generalized analysis.  Interestingly, when the graph is made directed, a novel class of instabilities reminiscent of the BF type can develop, for a choice of the parameter for which the latter are formally impeded on a symmetric support. To shed light onto this issue is entirely devoted the remaining part of this paper. We will in particular begin by considering a CGLE defined on a asymmetric lattice, with long range couplings. The adjacency matrix is in this case circulant, an observation that simplifies the mathematical characterization of the instability. In the second part of the paper we will turn to considering the case of an unspecified, directed and balanced, network and provide an alternative strategy to outline the conditions that seed the generalized BF instability.
\section{Periodic lattice with asymmetric long-range couplings}\label{circulant}
We begin by considering a specific class of directed regular lattices. These are periodic lattices, distorted with the inclusion of asymmetric long-ranged couplings. More specifically,  we will deal with a closed one dimensional ring composed of $N$ nodes. In a first place, we imagine each node to be linked to the $2 k$ neighbors, encountered when circulant the ring clockwise. We also consider self loops, and thus each node has an identical number of connections $M=2k+1$. Links are unweighted  but this latter assumption can be easily relaxed. The binary adjacency matrix which characterizes the introduced network is hence circulant, and balanced by definition. Starting from this setting, one can recursively apply a one-step shift operator to this latter adjacency matrix, to progressively reduce the imposed degree of asymmetry and eventually recover a symmetric network of inter-nodes connections.  Label with $n \leqslant k$  the positive integer which measures the number of repeated applications of the shift operator:  for $n=0$, the lattice 
is completely asymmetric, while the symmetry is restored for $n=k$.  As we will make clear in the following, understanding the class of instabilities that we here aim at describing amounts to characterize the spectrum of $\boldsymbol{\Delta}$, the Laplacian operator. For this reason, we will begin by characterizing the eigenvalues and eigenvectors of $\boldsymbol{\Delta}$, assuming the specific family of adjacency matrix outlined above.

A straightforward computation yields the following expression for the eigenvalues of $\boldsymbol{\Delta}$
\begin{equation}
\Lambda^{(\alpha)} = 1-M +\sum_{l=1}^{M-n-1} w_{\alpha}^l +\sum_{l=N-n}^{N-1} w_{\alpha}^l 
\end{equation}
and the associated eigenvectors read
\begin{equation}
\boldsymbol{\Phi}^{(\alpha)}=\frac{1}{\sqrt{N}} \left ( 1 , w_{\alpha}, w_{\alpha}^2,  w_{\alpha}^3, \ldots ,  w_{\alpha}^{N-1}  \right ) ^T
\end{equation}
where $ w_{\alpha}=\exp(2 \pi i \alpha /N)$ for $\alpha=0, \ldots, N-1$. 

It is important to remark that, for this specific class of networks, Eq. (\ref{eq:GLnetwork}) admits a family of solutions of the type:
\begin{equation}\label{eq:Weq}
{\bf W}^{TW}= \boldsymbol{\rho} e^{i \omega t}
\end{equation}
where   $\omega$ is real. Conversely, $\boldsymbol{\rho}$ is a complex quantity which satisfies the constraint:
\begin{equation}\label{eq:rho}
\boldsymbol{\rho}= \eta_{\beta}  \boldsymbol{\Phi}^{(\beta)}
\end{equation}
for all $\beta=0, \ldots, N-1$ and where we have omitted the index $\beta$ on the left hand side of the equation. The label $TW$ in Eq. (\ref{eq:Weq}) stands for traveling wave, as the above solutions share similarities with the traveling planar waves that manifest on a ordinary symmetric lattice with nearest neighbors couplings. 

Inserting Eq. (\ref{eq:Weq}) and (\ref{eq:rho}) into (\ref{eq:GLnetwork}), and recalling that $\vert \Phi^{(\beta)}_j \vert =1/\sqrt{N}$ $\forall \beta$, we get
\begin{equation}\label{eq:approxRho}
\vert \rho \vert ^2 \equiv \vert \rho_j \vert ^2 = \Lambda^{(\beta)}_{Re} - c_1  \Lambda^{(\beta)}_{Im} +1 
\end{equation}
and thus 
\begin{equation}\label{eq:approxOmega}
\omega = -c_2 \vert \rho \vert ^2 + c_1 \Lambda^{(\beta)}_{Re} + \Lambda^{(\beta)}_{Im} 
\end{equation}

It is worth emphasizing that Eq. (\ref{eq:Weq}) includes as a particular solution  the  homogeneous limit cycle $W^{LC} $ introduced above. This latter solution is in fact recovered for $\beta=0$ which gives  $\Lambda^{(0)}_{Re} = \Lambda^{(0)}_{Im}=0$ and, consequently, $\omega=-c_2$ and $\vert \rho \vert=1$.  

To proceed in the analysis we set down to investigate the stability of solutions (\ref{eq:Weq}). To this end we introduce a non homogeneous perturbation, which modifies the zeroth order solution $W^{TW}$ as:
\begin{equation}\label{eq:TW_SM_perturbed}
W_j=W^{TW}_j \left ( 1 + a_{+}(t) \Phi_j ^{(\alpha)}+ a_{-}(t) \bar{\Phi}_j^{(\alpha)}  \right )
\end{equation}
where the bar stands for the complex conjugate. The time dependent perturbations $a_{+}$ and $a_{-}$ are assumed small. By inserting Eq. (\ref{eq:Weq}) and Eq. (\ref{eq:TW_SM_perturbed}) into Eq. (\ref{eq:GLnetwork}), developing in the perturbation amount and  arresting the expansion at the linear order of approximation, one eventually gets:
\begin{equation*}
\frac{d}{dt}
\left (
\begin{matrix}
a_+\\
\bar{a}_-
\end{matrix}
\right)
=
{\bf J}
\left (
\begin{matrix}
a_+\\
\bar{a}_-
\end{matrix}
\right)
\end{equation*}
where the entries of matrix ${\bf J}$ read 
\begin{equation}\label{eq:matrixJ}
\begin{array}{l}
J_{11} = 1 -i \omega -2 (1+i c_2) \vert \rho \vert ^2 + (1 + i c_1) \Lambda^{(\alpha +\beta)} \\
J_{12} = \bar{J}_{21} = -(1+i c_2) \vert \rho \vert ^2 \\
J_{22} =  1 +i \omega -2 (1-i c_2) \vert \rho \vert ^2 + (1 - i c_1) \Lambda^{(\alpha-\beta)}.
\end{array}
\end{equation}
To carry out the calculation we made explicit use of the following relation:
\begin{equation}
\label{cond_circ}
\Phi_j ^{(\alpha)} \Phi_j ^{(\beta)}=\frac{1}{\sqrt{N}} \Phi_j ^{(mod(\alpha+\beta,N))}
\end{equation}
where $mod(\cdot, N)$ is the standard modulo $N$ operation. The above expression holds true because of the circulant nature of the Laplacian matrix. This latter condition cannot be invoked in general, a fact on which we shall return in the second part of the paper.  
The characteristic polynomial associated to matrix (\ref{eq:matrixJ})  yields the so called dispersion relation, a function of  $\Lambda^{(\alpha \pm \beta)} $ (the network analogues of the spatial wavelength), which ultimately determines the stability of $W^{TW}$. More specifically, compute the eigenvalues $\lambda^{(\alpha, \beta)}$ of matrix (\ref{eq:matrixJ}) and select the largest real part, $\lambda^{(\alpha, \beta)}_{Re}$. If $\lambda^{(\alpha, \beta)}_{Re}$ is positive for some $\Lambda^{(\alpha \pm \beta)}$, the TW solution is unstable to external non homogeneous  perturbation. 

The above general framework enables us to probe the stability of the limit cycle solution, the homogeneous configuration $W^{LC}$ characterized by a collection of  fully synchronized oscillators.  To this purpose set $\beta=0$ and denote with $\lambda^+(\alpha)=\lambda^{(\alpha, 0)}$ the eigenvalue of the matrix ${\bf J}$ with largest real part. By definition $\lambda^+(0)=\lambda^{(0, 0)} = 0$. To grasp the behavior of  $\lambda^+$ in the vicinity of $\alpha=0$ we perform a Taylor expansion of $\lambda^+(\alpha)$ in terms of  $\alpha/N$, treated as a continuum variable. A straightforward calculation returns: 
\begin{multline}
\lambda^+ \simeq 2 \pi  i  \left ( 1+c_1 c_2 \right  )  \left (k-n \right ) M\frac{ \alpha }{N} \\
+\left \{ -\frac{1}{6} M \left [ k \left (4 k +1 \right ) +3 n \left ( n-2 k \right ) \right ]  \left ( 1+c_1 c_2 \right  ) \right. \\
\left . \frac{1}{2} \left ( k-n \right )^2  M^2  c_1^2 \left ( 1 + c_2^2 \right )  \right \}  4 \pi^2 \frac{\alpha^2}{N^2}
 \end{multline}
The real part of $\lambda^+$ is therefore positive provided 
\begin{multline}\label{eq:stabilitySM}
 -\frac{1}{3} \left [ k \left (4 k +1 \right ) +3 n \left ( n-2 k \right ) \right ]  \left ( 1+c_1 c_2 \right  ) \\
+ \left ( k-n \right )^2  M c_1^2 \left ( 1 + c_2^2 \right ) >0
\end{multline}
When the two control parameters $(c_1,c_2)$ are chosen so as to match the above condition, the instability can develop and the fully synchronized state gets consequently disrupted. For any given choice of the pair ($c_1,c_2$), one can quantify from Eq. (\ref{eq:stabilitySM}) the degree of the imposed asymmetry (the smaller $n$ the more pronounced the asymmetry) that is to be enforced for the instability to develop.  Incidentally, we remark that Eq. (\ref{eq:stabilitySM}) constitutes the natural generalization of the homologous condition obtained in \cite{dipatti2016} for a CGLE with drift, on a continuous spatial support: the  
real quantity $\gamma$  therein defined (the ratio of the drift velocity and diffusion constants) is now formally replaced by a non trivial function of the model parameters $M$, $n$ and $k$.

To benchmark the above conclusion to the conventional framework, consider the limiting case of a symmetric matrix of connections, which amounts to selecting $n=k$ in (\ref{eq:stabilitySM}) to obtain:
\begin{equation}\label{eq:stabilitySM_old}
1+c_1 c_2 < 0
\end{equation}
namely the classical BF condition for the onset of the instability and its consequent synchronization loss. This latter condition is clearly more restrictive than the one derived for an asymmetric lattice of the type introduced above, for any choice of the shift parameter $n>1$. In practical terms, the imposed degree of asymmetry facilitates the dawning of a BF like instability, which can therefore develop in a region of the reference plane $(c_1,c_2)$ for which it is formally impeded in the symmetric setting. The usual BF instability is known to give rise to non homogeneous and intermittent, patchy density distributions. In \cite{dipatti2016} both TW solutions and heterogeneous patterns have been shown to occur in the extended region of BF instability, when operating on a continuum support, with a CGLE forced with a drift term. TW solutions are found when non homogeneous perturbation of a LC solution triggers a mode that is a stable attractor of the CGL dynamics. Patterns are conversely obtained when the selected mode is unstable:  the perturbation is no longer localized and spread to involve a, possibly large, collection of independent modes. What is going to happen when the system is instead hosted on a discrete asymmetric support? To answer to this question is entirely devoted the remaining part of this Section.

\begin{figure*}[tb]
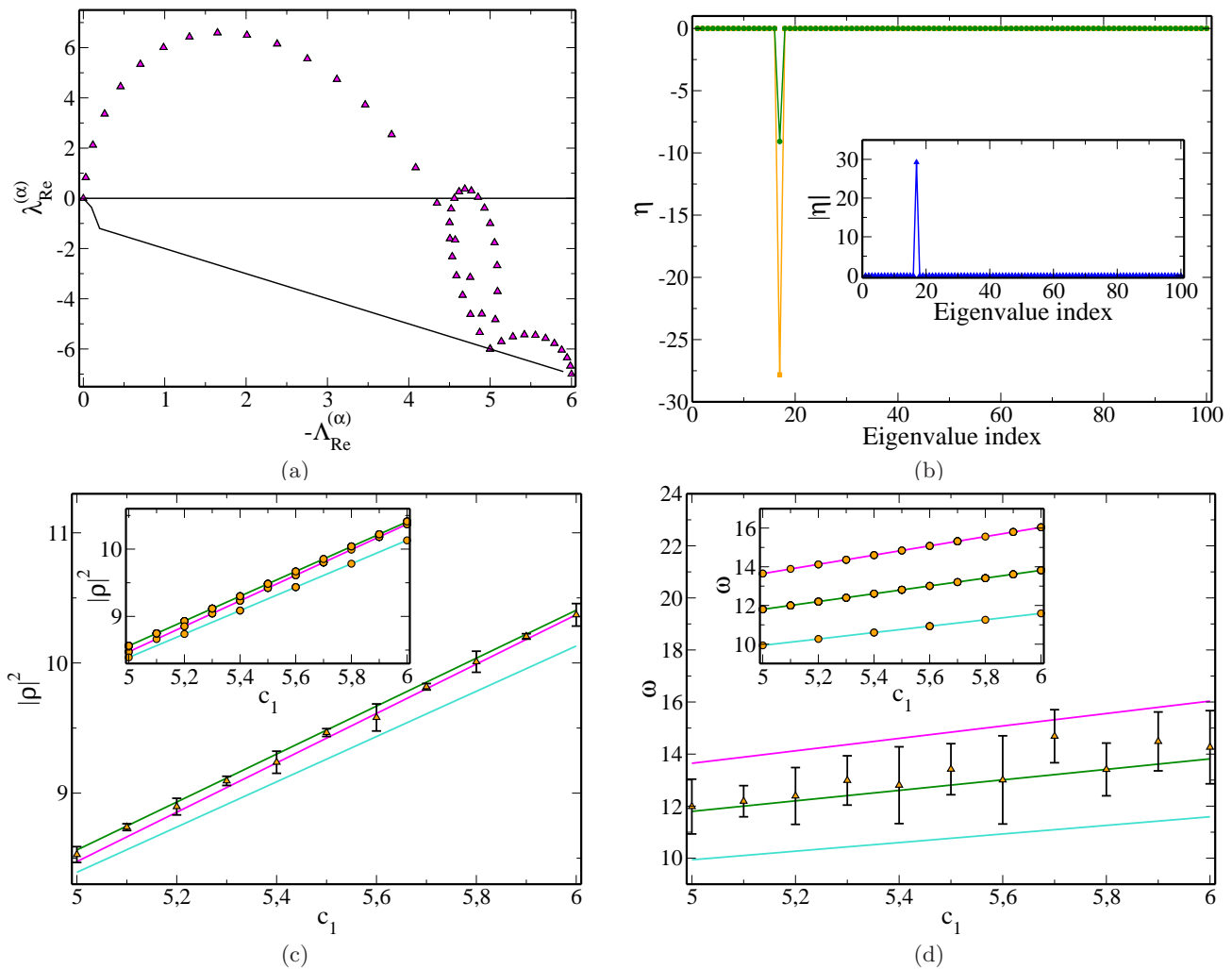

\begin{center}
\begin{tabular}{ccc}
\includegraphics[scale=0.3]{relDispShiftMatrix.eps} &
\phantom{ccc}&
\includegraphics[scale=0.3]{vetCShiftMatrix.eps}\\
(a) & \phantom{c}& (b)\\
\includegraphics[scale=0.3]{modQuadroVsc1_SM.eps} &
\phantom{ccc}&
\includegraphics[scale=0.3]{omegaVsc1_SM.eps}\\
(c) & \phantom{c}& (d)
\end{tabular}
\end{center}
\caption{Instability on an asymmetric circulant network  made of $N=100$ nodes and generated with $M=5$ and $n=1$. Panel (a): Real part of the dispersion relation $\lambda^{(\alpha, 0)}_{Re}\equiv \lambda^{(\alpha)}_{Re}$ as a function of  $-\Lambda^{(\alpha)}_{Re}$ (magenta (online) triangles). The maximum value of $\lambda^{(\alpha)}_{Re}$ is found at $\alpha_c=17$. The solid black line originates from the continuous theory. Main panel b): Real (orange (online) squares) and imaginary (green (online) dots)  components of the vector $\boldsymbol{\eta}$ obtained from one individual simulation of the CGLE. The LC synchronized state is initially perturbed with a tiny random disturbance. Inset of panel (b): modulus of $\boldsymbol{\eta}$. To numerically calculate $\boldsymbol{\eta}$ we have recorded the late time solution of  Eq. (\ref{eq:GLnetwork}), and multiplied the vector corresponding to the  last temporal iteration by the inverse of the matrix whose columns are formed by the eigenvectors ${\bf \Phi}^{(\alpha)}$ (see Appendix for further technical aspects). In panels (a) and (b)  parameters are set as  $c_1=5$ and  $c_2=0.2$. Panel (c): average amplitude perturbations as a function of $c_1$ ($c_2=0.2$ is kept fixed).  The solid lines correspond to the prediction given by Eq. (\ref{eq:rho}) for three different choices of the critical index $\alpha_c$ (turquoise  $\alpha_c=15$, green $\alpha_c=17$ and magenta $\alpha_c=19$). The (orange online) points in the inset originate from $10$ numerical integration of  Eq. (\ref{eq:GLnetwork}) with different initial realizations of the noise. The (orange online) triangles in the main plot represent the averaged computed quantities. The  
error bars correspond to one standard deviation. Panel (d) shows the comparison between the theoretical prediction of Eq. (\ref{eq:approxOmega}) and the data obtained by direct integration of Eq. (\ref{eq:GLnetwork}), as earlier described. Symbols and colors are chosen consistently with those described in the caption of panel (c).\label{fig:summaryShiftMatrix}}
\end{figure*}

To elaborate on this point, we assign the constants $c_1$ and $c_2$ so as to operate in the region of the asymmetry driven BF instability, as dictated by Eq. (\ref{eq:stabilitySM}). In Fig. \ref{fig:summaryShiftMatrix} (a) we display the dispersion relation $\lambda^{(\alpha, 0)}_{Re}\equiv \lambda^{(\alpha)}_{Re}$ as a function of  $-\Lambda^{(\alpha)}_{Re}$. The curve stays positive over a finite domain in $-\Lambda^{(\alpha)}_{Re}$, thus signaling the presence of an instability. In the linear regime of the evolution, the discrete modes that are made dynamically unstable, grow and take the system away from the initial synchronized configuration. For this specific choice of parameters, one can show that the most unstable mode, identified by the integer $\alpha_c$, is a stable stationary solution of the CGLE. Mathematically,  $\lambda^{(\alpha, \alpha_c)}_{Re}<0$, following the notation defined above. Similarly, the modes immediately adjacent to that associated to the global maximum of the dispersion relation, are also attractive equilibria of the CGLE. One can reasonably guess that the perturbation will eventually stabilize in a TW equilibrium of the type introduced in Eq. (\ref{eq:Weq}), with $\beta=\alpha_c$ as it happens for the continuum homologue system discussed in \cite{dipatti2016}. For the discrete setting, however, the dispersion relation is discontinuous and finite gaps separate adjacent active modes. There is therefore a  finite probability that the initially imposed random perturbation pushes the system towards the basin of attraction of a stationary TW solution (\ref{eq:Weq}) characterized by $\beta \ne \alpha_c$. We should also remark that the eigenvalues of the Laplacian operator come in conjugated complex pairs. Two distinct modes hide behind every symbol of  the dispersion relation profile, as depicted in Fig. \ref{fig:summaryShiftMatrix} (a). The mode that grows faster in the linear regime of the evolution is dynamically favored and hence outperforms the other.  This is the one that maximises $\vert \rho \vert ^2$ in Eq. (\ref{eq:approxRho}), as it follows from a straightforward calculation (not showed here).

To challenge the above interpretative picture we have carried out a campaign of simulations. The outcome of an individual realization is reported in Fig. \ref{fig:summaryShiftMatrix} (b) upon projecting the recorded signal on the basis formed by the eigenvectors of the Laplacian operator (see Appendix for further details). An isolated peak is evidently found,  confirming that the evolution self-consistently stabilizes one individual mode.  In panels (c) and (d)  of  Fig. \ref{fig:summaryShiftMatrix} we compare the measured value of $|\rho|^2$ and $\omega$ (symbols) vs. the theoretical predictions (\ref{eq:approxRho}) and (\ref{eq:approxOmega}) (solid lines), for different values of the control parameter $c_1$ (and for fixed $c_2$). The symbols are obtained by averaging the results obtained for different realizations of the imposed noise and the error bars testify on the degree of variability of the measure. The three solid lines, depicted with different colors, are drawn by considering the most unstable mode ($\alpha_c=17$), and the right and left nearest neighbors modes (rep. $\alpha= 15$ and $\alpha=19$). The symbols nicely populate the region that is enclosed between the two outer lines. More importantly, in the inset of panels (c) and (d)  of  Fig. \ref{fig:summaryShiftMatrix}, we report the result of each individual simulation without taking any ensemble average. The symbols fall exactly on top of one of the three depicted lines, so providing an a posteriori validation for the mechanism imagined above. The asymmetry enforced at the level of oscillators' couplings can destroy the LC synchrony, beyond the region of classical BF instability. Remarkably, the initial perturbation can eventually materialize in different TW solutions, which bear the indirect imprint of the linear dispersion relation in the vicinity of its global maximum.  

\begin{figure*}[tb]
\begin{center}
\includegraphics[scale=0.6]{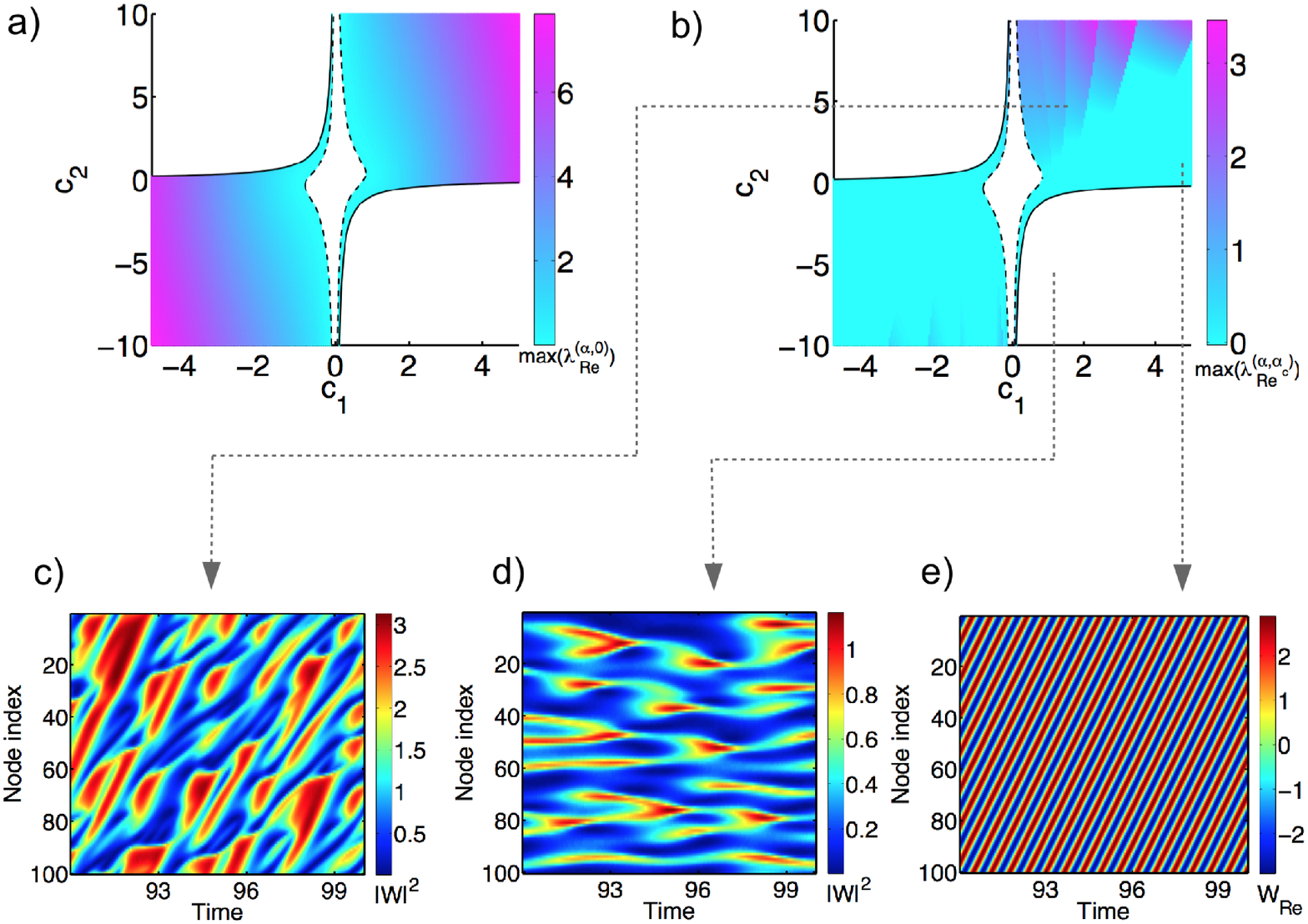}
\end{center}
\caption{Panel (a): the solid lines mark the transition to the BF instability, for an asymmetric networks of couplings. The dashed lines stem from condition (\ref{eq:stabilitySM}). The color refers to the maximum of the dispersion relation $\lambda^{(\alpha, 0)}_{Re}$. When positive $\lambda^{(\alpha, 0)}_{Re}>0$, the synchronized state is unstable to external perturbations. Panel (b):  $\lambda^{(\alpha, \alpha_c)}_{Re}$ is plotted with a proper color code, in the reference plan ($c_1,c_2$). Almost everywhere, $\lambda^{(\alpha, \alpha_c)}_{Re}>0$ signaling that the TW solution associated to $\alpha_c$ is generically unstable. Panels (c) and  (d) show the  evolution of the modulus of the complex amplitude $W_i(t)$ for, respectively, $(c_1, c_2)=(2,5)$ and $(c_1, c_2)=(1,-5)$. In panel (e) the real part of $W_i(t)$ is displayed for $(c_1, c_2)=(5,0.2)$. Patterns in panels (c) and (e) are obtained from direct numerical integration of Eq. (\ref{eq:GLnetwork}) on an asymmetric lattice of the type described in the main body of the paper, with $N=100$ nodes, $M=5$ and $n=1$. The pattern displayed in panel (d) is obtained for $n=2$ (symmetric long-ranged lattice). \label{fig:stabilitySM}}
\end{figure*}

Depending on the specific choice of the model parameters $c_1$ and $c_2$ more complex scenarios are however possible.  As it happens in the continuum limit \cite{dipatti2016}, the CGLE defined on an asymmetric discrete lattice can generate a rich gallery of patterns, reminiscent of the BF type. These patterns are however established outside the region deputed to the BF instability on a symmetric support, or, to state it differently, they occur for a choice of the parameters for which the synchronized configuration is classically predicted to be stable.  This result is illustrated via a composite collections of panels assembled in Fig.  \ref{fig:stabilitySM}, which refers to the specific setting $M=5$ and $n=1$.

\begin{figure}[tb]
\begin{center}
\includegraphics[scale=0.3]{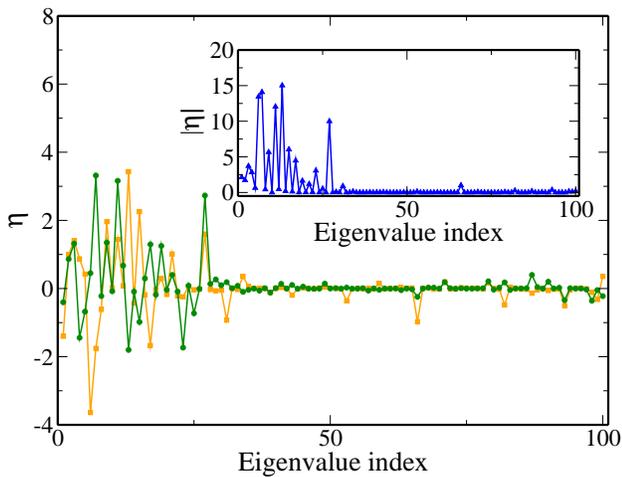}
\end{center}
\caption{Main panel: real (orange squares) and imaginary (green dots)  components of the vector $\boldsymbol{\eta}$, obtained by processing the late time pattern, as described in the main body of the paper. Inset: the modulus of $\boldsymbol{\eta}$ is plotted. Here, $c_1=2$ and $c_2=5$. Notice that this is the choice of the parameters that led to panel (c) in Fig. \ref{fig:stabilitySM}. 
\label{fig:eta_SM_pattern} }
\end{figure}

To further corroborate our speculative interpretation, we set down to analyze the pattern of panel (c). Following the protocol discussed above, we decompose the signal on the basis formed by the eigenvectors of matrix ${\bf \Delta}$. The results are plotted in Fig.  \ref{fig:eta_SM_pattern}: at variance with the case that lead to Fig. \ref{fig:summaryShiftMatrix} (b), several modes are simultaneously active an observation that contributes to explain the complex nature of the generated patterns. 

The dashed lines in panel (a) of Fig.  \ref{fig:stabilitySM} stand for the generalized condition of LC instability as specified by Eq. (\ref{eq:stabilitySM}). The solid lines mark the transition to the classical domain of BF instability, as obtained when operating with a symmetric web of connections. For all values of $(c_1,c_2)$ which fall inside the regions respectively delimited by the dashed and solid lines, non homogeneous perturbation magnifies and eventually destroys the perfect synchrony of the uniform limit cycle solution. The maximum of the dispersion relation $\lambda^{(\alpha, 0)}_{Re}$, when positive, is displayed with an apt color code, to confirm the theory predictions. We here solely scan the region $1+c_1 c_2 >0$, thus leaving aside the portion of the parameter plane which is customarily associated to the BF instability. Notice that the dashed lines accurately delineate the boundary of the region of interests, despite the approximations involved in the derivation of Eq. (\ref{eq:stabilitySM}). 

Recall that  $\alpha_c$ identifies location of the maximum of $\lambda^{(\alpha, 0)}_{Re}$. In other terms,  $\alpha_c$  refers to the most unstable mode, among those selected by  the inhomogeneous perturbation which is assumed to shake the synchronized LC configuration. Label with $\lambda^{(\alpha, \alpha_c)}_{Re}$ the maximum of the dispersion relation, that quantifies the stability of the TW solution associated to mode  $\boldsymbol{\Phi}^{(\alpha_c)}$. When  $\lambda^{(\alpha, \alpha_c)}_{Re} \le 0$, the mode that is spontaneously triggered following a modulational instability of the generalized BF type can eventually stabilize, in the non linear regime of the evolution. Conversely, when $\lambda^{(\alpha, \alpha_c)}_{Re}>0$, the instability interests a hierarchy of independent modes, possibly yielding structured patterns, the byproduct of the aforementioned unpredictable mixing. To validate this interpretative scenario, that we put forward building on the analogy with the discussion in  \cite{dipatti2016}, we report in panel (b) of Fig.  \ref{fig:stabilitySM} the quantity $\lambda^{(\alpha, \alpha_c)}_{Re}$, scanning the region of ($c_1,c_2$) where the asymmetry driven instability takes place. As one can readily appreciate, the TW solution that corresponds to the eigenvector $\boldsymbol{\Phi}^{(\alpha_c)}$ is stable close to the horizontal axis, hence for small values of $c_2$, which is, incidentally, the case that we have examined in Fig. \ref{fig:summaryShiftMatrix}. By performing a direct integration of the CGLE for a choice of the parameters that, according to the scheme outlined above, points to the TW stability, we obtain the asymptotic solution reported in panel (e), where the real part of $W_i$ is displayed. The simulation assumes an initial uniform LC state, modulated with a tiny non homogeneous perturbation. When the parameters ($c_1,c_2$) are instead selected so as  to have the system in the region with $\lambda^{(\alpha, \alpha_c)}_{Re}>0$, the modulus of the complex amplitude $W_j$ develops patchy patterns, in space and time, as shown in panel (c) of Fig. \ref{fig:stabilitySM}.  These latter patterns closely resemble those obtained in the region conventionally  deputed to the BF instability: to help comparison, in panel (d) we report the pattern obtained for a {\it symmetric} ($n=2$, $M=5$) lattice of connections between oscillators.

Starting from this setting, it would be interesting to extend the analysis beyond the case here considered, namely by relaxing the hypothesis of dealing with a circulant matrix of connections. This latter modification yields important consequences.  First of all,  the quantity $|{\Phi}^{(\alpha)}_j|$, the modulus of each individual component $j$ of the Laplacian eigenvector ${\bf \Phi}^{(\alpha)}$, is not constant, but, instead, depends on the indexes $j$ and $\alpha$.   Hence, the cubic term in the CGLE cannot be simply managed, as opposed to what it happens for the simplified case examined in this Section. Consequently,  the TW ansatz (\ref{eq:Weq}), complemented with Eq. (\ref{eq:rho}), cannot be invoked as a prototypical solution for the general formulation of the problem. If the network of connections is however balanced (identical incoming and outgoing connectivity, per node), the homogeneous limit cycle $W_j \equiv W^{LC}$ is still a solution of the CGLE.  We can therefore perform a linear stability analysis around such an homogeneous, time dependent, solution to elaborate on the conditions that result in a loss of the initial synchronization, as follows the introduction of a non homogeneous disturbance. To address this point is entirely devoted the forthcoming Section.

\section{The case of a generic balanced and directed network}\label{network}
As anticipated above, we now turn to considering the case of a  CGLE defined on a directed, heterogeneous although balanced, network. Since the network of mutual connections is balanced, the uniform limit cycle, a fully synchronized replica of the periodic orbit of the a-spatial system, solves the spatially extended CGLE. We are here interested in isolating the parameters condition that yields a loss of synchronization, when the uniform state is altered with the injection of a non homogeneous perturbation. To carry out the linear stability analysis, we shall follow an apt  procedure, inspired to those reported in \cite{nakaoCL, nakao2014}. We will in particular schematize the imposed perturbation as:
\begin{equation*}
W_j (t) =W^{LC}(t) (1 + \nu_j(t))e^{i \theta_j( t)}
\end{equation*}
and insert the above ansatz into Eq. (\ref{eq:GLnetwork}). By linearizing around the uniform solution $\nu_j=0$ and $\theta_j=0$,  one eventually gets: 
\begin{equation}\label{eq:linearization}
\left (
\begin{matrix}
\dot{\nu}_j\\
\dot{\theta}_j
\end{matrix}
\right)
= 
\left (
\begin{matrix}
-2 & 0 \\
-2 c_2 & 0
\end{matrix}
\right)
\left (
 \begin{matrix}
\nu_j\\
\theta_j
\end{matrix}
\right)
+ \sum_{k=1}^N \Delta_{jk}
 \left (
\begin{matrix}
1 & -c_1 \\
c_1 & 1
\end{matrix}
\right)
\left (
 \begin{matrix}
\nu_k\\
\theta_k
\end{matrix}
\right) \quad .
\end{equation}

To proceed with the analysis, we introduce the eigenvectors ${\bf \Phi}^{(\alpha)}$ of the Laplacian operator ${\bf \Delta }$. In formulae  ${\bf \Delta } {\bf \Phi}^{(\alpha)} = \Lambda^{(\alpha)}{ \bf \Phi}^{(\alpha)}$. We further assume that the eigenvectors ${\bf \Phi}^{(\alpha)}$ are linearly independent. This is not always the case when the underlying graph is directed. For this reason, the diagonalizability of the Laplacian matrix is a minimal requirement that needs to be satisfied for the following analytical treatment to hold true. Under this working assumption,  one can expand the perturbations $\nu_j$ and $\theta_j$ as:
\begin{equation}\label{eq:expansionEigenvectorsDelta}
\begin{pmatrix}
\nu_j \\ \theta_j
\end{pmatrix}= \sum_{\alpha=1}^N
\begin{pmatrix}
\nu^{(\alpha)} \\ \theta^{(\alpha)}
\end{pmatrix}
e^{\lambda^{(\alpha)} t} \Phi_j^{(\alpha)}
\end{equation}
where $\nu^{(\alpha)} $  and $\theta^{(\alpha)}$ denote the expansion coefficients, while $\lambda^{(\alpha)}$ sets the characteristic growth rate of the $\alpha$-th mode. Plugging expressions (\ref{eq:expansionEigenvectorsDelta}) into Eq. (\ref{eq:linearization}) yields  the following condition: 
\begin{equation}
\label{eq:disp_rel}
\text{det} 
\left(
\begin{matrix}
-2 + \Lambda^{(\alpha)}-\lambda^{(\alpha)} & - c_1 \Lambda^{(\alpha)}\\
-2 c_2 + c_1 \Lambda^{(\alpha)} & \Lambda^{(\alpha)}- \lambda^{(\alpha)}
\end{matrix}
\right) = 0 
\end{equation}
that should be matched by $\lambda^{(\alpha)}$, the complex quantity which sets the rate of the exponential growth of the perturbation. Equivalently, in a more compact notation,  we require $\text{det}  \left (  {\bf J}_{\alpha}  - \lambda^{(\alpha)} \mathbb{I}_{2}  \right  ) =0$ with:
\begin{equation*}
{\bf J}_{\alpha} = \left (
\begin{matrix}
-2 +\Lambda^{(\alpha)} & -c_1 \Lambda^{(\alpha)}\\
-2 c_2 + c_1 \Lambda^{(\alpha)} & \Lambda^{(\alpha)}
\end{matrix}
\right ) \qquad .
\end{equation*}
Recall that  $\Lambda^{(\alpha)}$ is a complex quantity, as the underlying network is assumed directed, and the associated Laplacian consequently asymmetric. Furthermore, we remind that  $\Lambda_{Re}^{(\alpha)} <0$, since the spectrum of the Laplacian matrix  falls in the left half of the complex plane, according to the Gerschgorin circle theorem \cite{gersh}. A simple algebraic manipulation yields therefore the following relation, with an obvious meaning of the adopted notation:
\begin{equation}\label{eq:det_trace}
\begin{aligned}
(\text{tr} {\bf J}_{\alpha})_{Re}  & =  -2 + 2 \Lambda_{Re}^{(\alpha)}\\
(\text{tr} {\bf J}_{\alpha})_{Im} & =   2 \Lambda_{Im}^{(\alpha)} \\
(\text{det} {\bf J}_{\alpha})_{Re}  & =  - 2 \Lambda_{Re}^{(\alpha)} +  (\Lambda_{Re}^{(\alpha)})^2 -   (\Lambda_{Im}^{(\alpha)})^2 \\
&\phantom{ = } -2 c_1 c_2   \Lambda_{Re}^{(\alpha)} + c_1^2 \left [ (\Lambda_{Re}^{(\alpha)})^2 -  (\Lambda_{Im}^{(\alpha)})^2 \right  ]\\
(\text{det} {\bf J}_{\alpha})_{Im}  & =  -2   \Lambda_{Im}^{(\alpha)} + 2 (1 + c_1^2)\Lambda_{Re}^{(\alpha)} \Lambda_{Im}^{(\alpha)} \\
&\phantom{ = } - 2 c_1 c_2 \Lambda_{Im}^{(\alpha)}  \Lambda_{Im}^{(\alpha)}
\end{aligned}
\end{equation}
The quantity $\lambda^{(\alpha)}$, stemming from Eq. (\ref{eq:disp_rel}), writes therefore:
\begin{equation}
\lambda^{(\alpha)}=\frac{1}{2}\left[(\text{tr} {\bf J}_{\alpha})_{Re}+\gamma \right]+\frac{1}{2}\left[(\text{tr} {\bf J}_{\alpha})_{Im}+\delta \right] i
\end{equation}
where:
\begin{equation}
\gamma=\sqrt{\frac{a+\sqrt{a^2+b^2}}{2}}
\end{equation}
\begin{equation}
\delta=sgn(b) \sqrt{\frac{-a+\sqrt{a^2+b^2}}{2}}
\end{equation}
and:
\begin{equation}\label{eq:a_b}
\begin{aligned}
a &=[(\text{tr} {\bf J}_{\alpha})_{Re}]^2 - [(\text{tr} {\bf J}_{\alpha})_{Im}]^2 - 4 (\text{det} {\bf J}_{\alpha})_{Re}\\
b &= 2 (\text{tr} {\bf J}_{\alpha})_{Re} (\text{tr} {\bf J}_{\alpha})_{Im} - 4 (\text{det} {\bf J}_{\alpha})_{Im}
\end{aligned}
\end{equation}

As emphasized in \cite{asllani14},  diffusion driven instabilities can  develop also when $\text{tr}({\bf J}_{\alpha})_{Re}<0$, at variance with what it happens when the inspected system is hosted on a symmetric spatial support. In fact,  $\lambda_{Re}^{(\alpha)}>0$ if:
\begin{equation}\label{eq:inequality}
\vert (\text{tr} {\bf J}_{\alpha})_{Re}  \vert \leqslant  \sqrt{\frac{a+\sqrt{a^2+b^2}}{2}}
\end{equation}
a condition that can be met for $\text{tr} ({\bf J}_{\alpha})_{Re}<0$, provided the symmetry of the couplings is broken and an imaginary component of the Laplacian spectrum is hence active. For this reason, this family of instabilities has been referred to as to topology driven \cite{asllani14}. Indeed, the aim of this calculation is to generalize the analysis of \cite{asllani14} to the setting where the unperturbed homogeneous solution is a time dependent limit cycle and not just a trivial fixed point. This scenario is here exemplified by the CGLE. 

A straightforward, although lengthy, calculation  allows one to recast the condition for the instability in the form:
\begin{equation}\label{eq:stabilityReteGen}
S_2(\Lambda_{Re}^{(\alpha)}) \leqslant S_1(\Lambda_{Re}^{(\alpha)}) \left [ \Lambda_{Im}^{(\alpha)} \right ] ^2
\end{equation}
where 
\begin{equation}
\begin{aligned}
S_2(\Lambda_{Re}^{\alpha}) &=C_{2,4}(\Lambda_{Re}^{(\alpha)})^4-C_{2,3}(\Lambda_{Re}^{(\alpha)})^3+C_{2,2}(\Lambda_{Re}^{(\alpha)})^2\\
& \phantom{= }-C_{2,1}\Lambda_{Re}^{(\alpha)} \\
S_1(\Lambda_{Re}^{\alpha}) &=C_{1,2}(\Lambda_{Re}^{\alpha})^2 -C_{1,1}\Lambda_{Re}^{(\alpha)} +C_{1,0}
\end{aligned}
\end{equation}
with
\begin{equation}
\label{eq:coeff}
\begin{aligned}
C_{2,4}&=1+c_{1}^2  \\
C_{2,3}&=4+2c_{1}c_{2}+2c_{1}^2 \\
C_{2,2}&=5+4c_{1}c_2+c_{1}^2 \\
C_{2,1}&=2+2c_{1}c_2 \\
C_{1,2}&=c_1^4+c_1^2 \\
C_{1,1}&=2c_1^{3}c_2+2c_1^2  \\
C_{1,0}&=c_{1}^{2}(1+c_2^2) \qquad .
\end{aligned}
\end{equation}

As an obvious consistency check, we remark that Eq. (\ref{eq:stabilityReteGen}) reduces to $S_2(\Lambda_{Re}^{(\alpha)})<0$ when dropping the imaginary components of  $\Lambda^{(\alpha)}$, or, equivalently, when  dealing with a symmetric network of couplings. It is immediate to realize that such a limiting inequality yields the standard BF condition, namely $1+c_1 c_2 <0$.

The remaining part of this Section is devoted to elaborate on the implications of the above analysis. To this end we generate a directed network, via a suitable  modification of the Newman-Watts (NW) algorithm. More specifically, we start from a substrate $K$-regular ring made of $N$ nodes. We then add, on average, $NKp$ long-range directed links, in addition to the links due to the regular lattice. Here, $p \in [0, 1]$ is a probability that can be chosen by the user and which quantifies the amount of long-ranged distortion accommodated for.  
To keep the network balanced (identical number of incoming and outgoing links, per node), the inclusion of a long-range link originating from node $j$ is followed by the insertion of a fixed number ($3$ is our arbitrary choice) of additional links to form a loop that closes on $j$  \cite{asllani14}.

\begin{figure*}[tb]
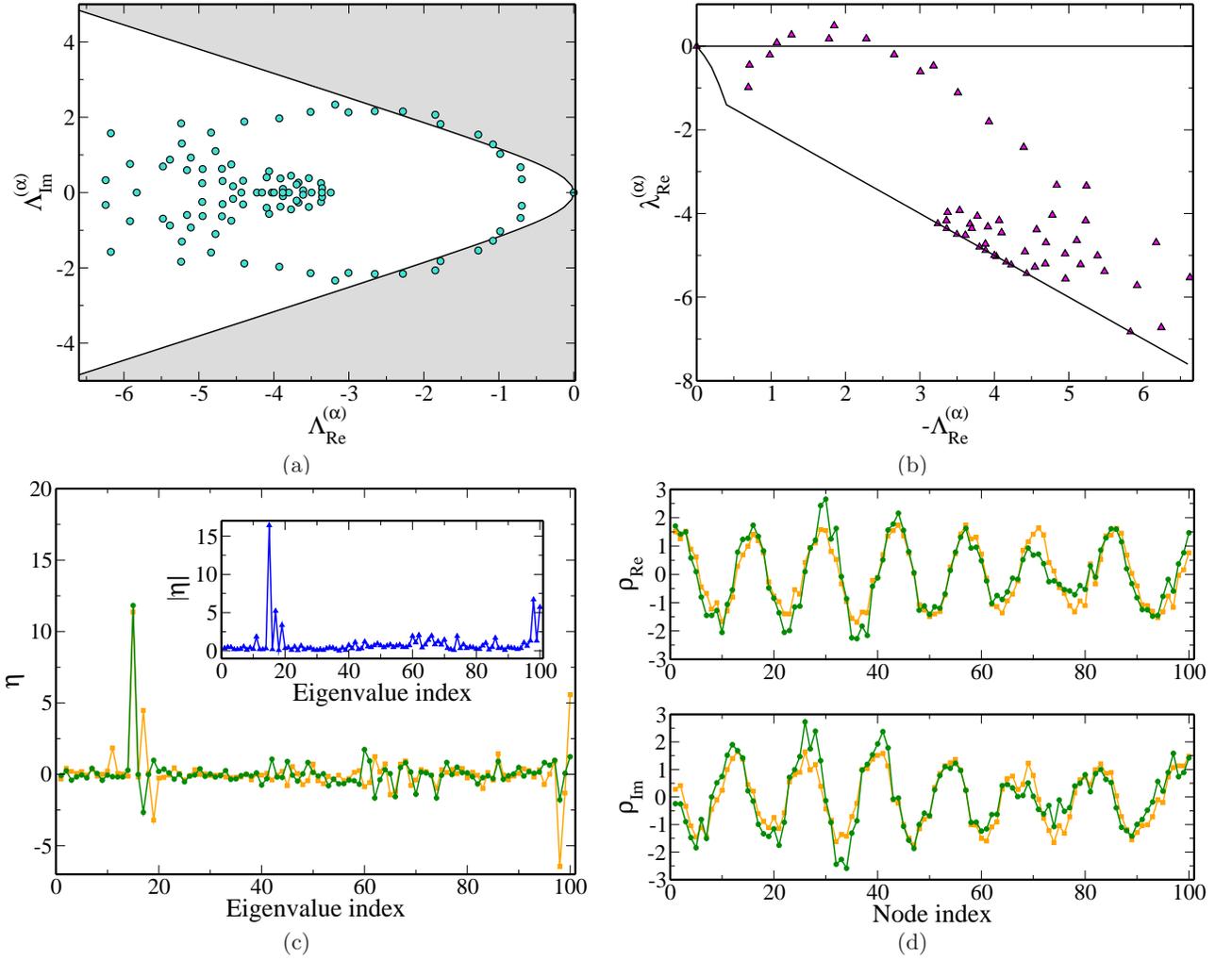

\begin{tabular}{ccc}
\includegraphics[scale=0.3]{stabilityRegionReteGen.eps} &
\phantom{c}&
\includegraphics[scale=0.3]{relDisp_c1_1.564.eps}\\
(a) & \phantom{c}& (b)\\
\includegraphics[scale=0.3]{vetC_c1_1.564.eps}
&
\phantom{c}&
\includegraphics[scale=0.3]{solApproxMatrixGen_c1_1.564.eps}\\
(c) & \phantom{c}& (d)
\end{tabular}
\caption{Instability on a network made of $N=100$ nodes and  generated with a modified NW algorithm with $p=0.27$. Parameters are set as $c_1=1.564$ and $c_2=0.675$. Panel (a): Instability region (shaded area) predicted by Eq. (\ref{eq:inequality}) and depicted in the plan $(\Lambda_{Re}^{(\alpha)},\Lambda_{Im}^{(\alpha)} )$. Cyan dots stand for the eigenvalues of the Laplacian matrix. Panel (b): the (magenta online) triangles  show the dispersion relation given by Eq. (\ref{eq:disp_rel}). The black solid lines originate from the symmetric (and continuous) counterpart.  Panel (c):  the late time pattern, stemming from the instability, projected on the basis formed by the eigenvectors of the Laplacian operator (see Appendix for details on the procedure). Real  (orange (online) squares), the imaginary (green (online) circles) and the modulus (blue (online) triangles) of the vector $\boldsymbol{\eta}$ are displayed.  Panels (d): comparison between (i) the real and imaginary components of $\boldsymbol{W}$ -- as obtained via direct numerical integration of Eq. (\ref{eq:GLnetwork}) -- (orange (online) squares) and (ii) an the approximate solution (\ref{ansatz}) (green (online) circles) which uniquely account for the most unstable mode, as identified by dispersion relation relation analysis. 
\label{fig:genericNet}}
\end{figure*}

The generalized condition  (\ref{eq:stabilityReteGen}) for the instability on a directed (balanced) network can be graphically illustrated in the reference plane $(\Lambda_{Re}^{(\alpha)},\Lambda_{Im}^{(\alpha)} )$. Given the reaction parameters $c_1$ and $c_2$, one can determine the coefficients $C_{1,q} (q=0, ..., 4)$ and $C_{2,q} (q=0, 1, 2)$ via the above Eqs. (\ref{eq:coeff}). Inequality (\ref{eq:stabilityReteGen}) enables us to delimit a region of instability, represented as a shaded domain in Fig.  \ref{fig:genericNet}(a). Each eigenvalue of the discrete Laplacian appears as a localized point in the plan $(\Lambda_{Re}^{(\alpha)},\Lambda_{Im}^{(\alpha)} )$. The instability takes place if a subset of the $N$ eigenvalues invade the shaded region isolated above. The asymmetry acts as a crucial ingredient: for an undirected graph,  the points lay on the real (horizontal) axis, thus outside the predicted domain of instability. In Fig.  \ref{fig:genericNet}(b) the dispersion relation $\lambda_{Re}^{(\alpha)}$ is plotted as function of $-\Lambda_{Re}^{(\alpha)}$. The black solid line refers to limiting case of a symmetric (and continuum) support.  No instability can develop, when the symmetry is enforced into the system, as, in this case,  $\lambda_{Re}^{(\alpha)}<0$. The discrete collection of points shown in  Fig.  \ref{fig:genericNet}(a) refers to the directed case, namely the situation that led to Fig.  \ref{fig:genericNet}(a). The points lift above the solid curve and cross the horizontal axis, thus signaling the presence of a topology driven instability of the uniform limit cycle solution. As usual, the instability will eventually give rise to complex patterns, in the non linear regime of the evolution. In the following we will seek to gain insight into the specific traits of the self-emerging patterns. 

As a preliminary observation, we notice that the asymptotic signal obtained upon integration of the CGLE displays a rather simple structure in time. Label $W_j(t) $ the recorded amplitude on node $j$ at time $t$, after the initial transient has been filtered out. By performing a conventional discrete Fourier transform in time, one can convincingly conclude that  $W_j(t) = \rho_j e^{i \omega t}$ where $\rho_j$ is a constant (time independent) non homogeneous (node dependent) rescaled amplitude factor. In other words the emerging pattern can be seen as a discrete collection of periodic oscillators, characterized by different amplitudes ($\rho_j$) and identical periods ($2 \pi /\omega$). Given these facts, is it possible to obtain an approximated (closed analytical) solution of the CGLE  that matches to some extent the results of the simulation? The difficulty stems from the non linear (cubic) term in the CGLE. As already emphasized, $|{\Phi}^{(\alpha)}_j|$ is not constant over $j$ and for this reason ansatz (\ref{eq:Weq}) -- together with (\ref{eq:rho}) -- cannot be swiftly invoked.  On the other hand, it seems plausible to hypothesize that the asymptotic pattern bears the imprint of the dispersion relation, which holds in the linear regime of the evolution. More specifically it could be argued that the spatial properties of the emerging pattern are, at least  approximately, explained in terms of ${\bf \Phi}^{(\alpha_c)}$, the eigenvector associated to the most unstable mode, as identified via the linear stability analysis. It is therefore tempting to set:
\begin{equation}
\label{ansatz}
{\bf W}(t) \simeq \rho^{(\alpha_c)} {\bf \Phi}^{(\alpha_c)} e^{i \omega_{\alpha_c} t}
\end{equation}
where $\rho^{(\alpha_c)}$ and $ \omega_{\alpha_c} $ are quantities that need to be self-consistently determined.  To this end, we propose to approximate the term $| W_j|^2$ in the CGLE with an constant expression which follows ansatz (\ref{ansatz}):
\begin{equation}
\label{ansatz1}
|W_j|^2 \simeq |\rho^{(\alpha_c)}|^2 \langle |{\bf \Phi}^{(\alpha_c)} |^2 \rangle \equiv   |\rho^{(\alpha_c)}|^2  \frac{1}{N} \sum_k^{N}  |{\Phi_k}^{(\alpha_c)} |^2
\end{equation}
Importing the above approximation in the CGLE, yields a linear equation in ${\bf W}$. Importantly, relation (\ref{ansatz}) solves the obtained equation provided $|\rho^{(\alpha_c)}|$ and $\omega_{\alpha_c}$ match the following self-consistent conditions:
\begin{eqnarray*}
\omega_{\alpha_c}+c_2 |\rho^{(\alpha_c)}|^2 \langle |{\bf \Phi}^{(\alpha_c)} |^2 \rangle-\Lambda_{Im}^{(\alpha_c)}-c_1 \Lambda_{Re}^{(\alpha_c)} &=& 0 \\
-1+ |\rho^{(\alpha_c)}|^2  \langle |{\bf \Phi}^{(\alpha_c)} |^2 \rangle - \Lambda_{Re}^{(\alpha_c)}+c_1 \Lambda_{Im}^{(\alpha_c)} &=& 0
 \end{eqnarray*}
We are therefore in a position to analytically estimate the unknown quantities $\rho^{\alpha_c}$ (up to a unimportant phase constant) and $\omega_{\alpha_c}$ and so to challenge the approximate closed form solution (\ref{ansatz}) versus direct simulations. The comparison is performed in Fig. \ref{fig:genericNet}(d) and returns a satisfying agreement. Only a limited fraction of modes contributes indeed to the recorded signal.  
This is manifestly evident in Fig.  \ref{fig:genericNet}(c): here, the  late time pattern is projected on the basis of the eigenvectors of the Laplacian operator, following the procedure described in the Appendix. This observation motivates a posteriori the choice of having proposed the closed approximate solution (\ref{ansatz}). To build an ideal bridge with the analysis carried out in the  preceding section, we can interpret the pattern obtained above as an effective quasi-traveling wave profile which localizes on the most unstable mode, as triggered by the imposed perturbation. Also in this case, more intricate situations can be generated when the set of unstable modes gets enlarged, in complete analogy with what it happens  for the simpler setting analyzed in the preceding Section. 

\section{Conclusions}
The spontaneous ability of spatially extended system to self-organize in space and time is proverbial and has been raised to a paradigm in several fields of investigations. Under specific conditions, trivial homogeneous solutions get destroyed by the injection of a tiny, non homogeneous perturbation. This latter can in fact grow exponentially, in the linear regime of the evolution, following an instability mechanism that originates from the complex interplay between the reaction and diffusion (or, more generally, spatial) components of the dynamics. In this paper we have focused on the celebrated Benjamin-Feir instability within the framework of the Complex Ginzburg-Landau equation (CGLE), a reference model for non linear studies. In  \cite{dipatti2016} it was shown that generalized BF instabilities can set in, instigated by a drift term,  outside the region of parameters for which the conventional BF instability manifests. This latter study assumed a continuum, one dimensional, spatial support. In this paper, we have taken one step forward by assuming that the population of discrete non linear oscillators is linked through a directed network of connections. As we here proved, directionality drives a degree of anisotropy, conceptually similar to drift in the continuum representation, which can trigger the sought instability.

To reach this conclusion we have preliminary considered a rather specific class of networks, characterized by a circulant adjacency matrix.  Perturbed homogeneous limit cycles, can give rise to TWs or patchy intermittent patterns, these latter being remarkably similar to those displayed within the classical BF domain. We then proceeded to consider the case of  a generic, though balanced, directed graph. The conditions that yields the asymmetry driven BF instability are obtained by adapting the analysis of \cite{asllani14} to the current setting where the homogeneous solution depends explicitly on time.  An approximate closed analytical solution is proposed which enables one to mimic the patterns recorded, via numerical integration of the CGLE. Overall we have contributed to the current understanding of topology driven instability, elaborating further on the crucial role played by the network specificity in shaping the, spontaneously emerging, spatially extended patterns. 
\section*{Appendix}
Let us denote by $\boldsymbol{ \mathcal{F}}$ the $N \times N$ matrix whose columns consist of the eigenvectors $\boldsymbol{\Phi}^{(\alpha)}$  of the Laplacian matrix $\boldsymbol{\Delta}$.  Since the eigenvectors form a complete basis, we can express the amplitude vector $\boldsymbol {\rho}$ (see Eq. (\ref{eq:Weq}))  in terms of this novel basis,  through the following change of variables
\begin{equation}
\boldsymbol {\rho} = \boldsymbol{ \mathcal{F}} \boldsymbol {\eta}
\end{equation} 
where  $\boldsymbol {\eta}$ is the vector of the new coordinates.  

To compute $\boldsymbol {\eta}$ numerically we proceed as follows. First, we numerically integrate the CGLE and then  record the late time solution $\bf W$. Then, we estimate numerically the frequency $\omega$, by processing the registered signal $\bf W$ via a  standard time discrete Fourier Transform. By recalling that  $\boldsymbol{ \mathcal{F}}$  is invertible,  one gets:  
\begin{equation}
\boldsymbol {\eta} = \boldsymbol{ \mathcal{F}}^{-1} \bf W e^{-i \omega t} \quad .
\end{equation}
\begin{acknowledgments}
This work has been supported by the program PRIN 2012 founded by the Italian Ministero dell'Istruzione, dell'Universit\`{a} e della Ricerca (MIUR). D.F. acknowledges financial
support from H2020-MSCA-ITN-2015 project COSMOS 642563.  The work of T.C. presents research results of the Belgian Network DYSCO (Dynamical Systems, Control, and Optimization), funded by the Interuniversity Attraction Poles Programme, initiated by the Belgian State, Science Policy Office.
\end{acknowledgments}
\bibliography{bibliography}{}
\end{document}